\documentclass[11pt]{article}
\usepackage[dvips]{graphicx}
\usepackage{mathrsfs}
\usepackage{amsthm} 
\usepackage{amsmath}
\usepackage{amssymb}
\usepackage{amsfonts}
\usepackage{float}
\usepackage{yhmath}
\usepackage{wrapfig}
\usepackage[dvips]{color}
\usepackage{cite}
\usepackage{array}
\usepackage{tabularx}
\usepackage[sectionbib]{chapterbib}
\usepackage{threeparttable}
\usepackage{chemarrow}
\usepackage{framed}
\definecolor{shadecolor}{gray}{0.90}
\usepackage[small, bf, labelsep=quad]{caption}
\newcommand{\BL}[1]{\textcolor{#1}{$-\hspace{-0.6mm}-$}}

\begin{document}

\begin{center}
\Huge{Segment Distribution around the Center of Gravity}
\end{center}
\vspace*{-6mm}
\begin{center}
\Huge{of a Triangular Polymer}
\end{center}

\vspace*{5mm}
\begin{center}
\large{Kazumi Suematsu\footnote{\, The author takes full responsibility for this article.}, Haruo Ogura$^{\dagger 2}$, Seiichi Inayama$^{\dagger 3}$, and Toshihiko Okamoto$^{\dagger 4}$} \vspace*{2mm}\\
\normalsize{\setlength{\baselineskip}{12pt} 
$^{\dagger 1}$ Institute of Mathematical Science\\
Ohkadai 2-31-9, Yokkaichi, Mie 512-1216, JAPAN\\
E-Mail: suematsu@m3.cty-net.ne.jp, ksuematsu@icloud.com  Tel/Fax: +81 (0) 593 26 8052}\\[3mm]
$^{\dagger 2}$ Kitasato University,\,\, $^{\dagger 3}$ Keio University,\,\, $^{\dagger 4}$ Tokyo University\\[15mm]
\end{center}

%%%%%%%%%%%%%%%%%%
\hrule
\vspace{3mm}
\noindent
\textbf{\large Abstract}: The segment distribution around the center of gravity is investigated for a special comb polymer (triangular polymer) having the side chains of the same generation number, $g$, as the main backbone. Common to all the other polymers, the radial mass distribution is expressed as the sum of the distribution functions for the end-to-end vectors, $\{\vec{r}_{Gh}\}$, from the center of gravity to the monomers on the $h$th generation; the result being, for a large $g$, 
%%%%%%%%%%%%%%%%%%
\begin{multline}
\varphi_{\text{triang}}(s)=\frac{1}{N}\left\{\sum_{h=1}^{g}\left(\frac{d}{2\pi\left\langle r_{Gh}^{2}\right\rangle}\right)^{\frac{d}{2}}\text{Exp}\left(-\frac{d}{2\left\langle r_{Gh}^{2}\right\rangle}s^2\right)\right.\\
\left.+\sum_{h=2}^{g}\sum_{j=1}^{g-h}\left(\frac{d}{2\pi\left\langle r_{Gh_{j}}^{2}\right\rangle}\right)^{\frac{d}{2}}\text{Exp}\left(-\frac{d}{2\left\langle r_{Gh_{j}}^{2}\right\rangle}s^2\right)\right\}\notag
\end{multline}
It is found that the mean square of the radius of gyration varies as $\left\langle s_{N}^{2}\right\rangle_{0}\doteq\frac{7}{15}\,g\,l^{2}$, as $g\rightarrow\infty$. Since $g\propto \sqrt{N}$ for the triangular polymer, this leads to $\left\langle s_{N}^{2}\right\rangle_{0}^{1/2}\propto N^{1/4}$, giving the same exponent as observed for the randomly branched polymer. On the basis of the present result, we put forth that all the known polymers obey the equality: $\left\langle s_{N}^{2}\right\rangle_{0}=A\, g\,l^{2}$, where $A$ is a polymer-species-dependent coefficient and also depends on the choice of the root monomer. We discuss the extension of this empirical equation.
\vspace{-2mm}

\begin{flushleft}
\textbf{\textbf{Key Words}}: Segment Distribution/ Triangular Polymer/ Comb Polymer/ Exponent $\nu_{0}$/
\normalsize{}\\[3mm]
\end{flushleft}
\hrule
\vspace{3mm}
\setlength{\baselineskip}{13pt}

%%%%%%%%%%%%%%%%%% Introduction
\section{Introduction}
In the preceding work\cite{Kazumi}, we discussed that a randomly branched polymer is a mixture of a variety of isomers: a linear polymer, star polymers, irregularly branched polymers, and dendrimers. One way to classify such diverse isomers is to characterize the spacial configurations according to the asymptotic relation, $\left\langle s_{N}^{2}\right\rangle_{0}\propto N^{2\,\nu_{0}}$ for $N\rightarrow\infty$, between the radius of gyration and the total mass. As has been well-established, an unperturbed randomly-branched-polymer has the exponent, $\nu_{0}=\tfrac{1}{4}$, whereas the linear, star, and comb polymers have the exponent $\nu_{0}=\tfrac{1}{2}$, and the dendrimers have $\nu_{0}=0$. A randomly branched polymer has just the intermediate exponent between $1/2$ and 0. To date, within our knowledge, no single branched polymer having $0<\nu_{0}<1/2$ has been reported. In this study, we focus on a special comb polymer with side chains comparable in length to the backbone.

\begin{figure}[h]
\begin{center}
\includegraphics[width=17cm]{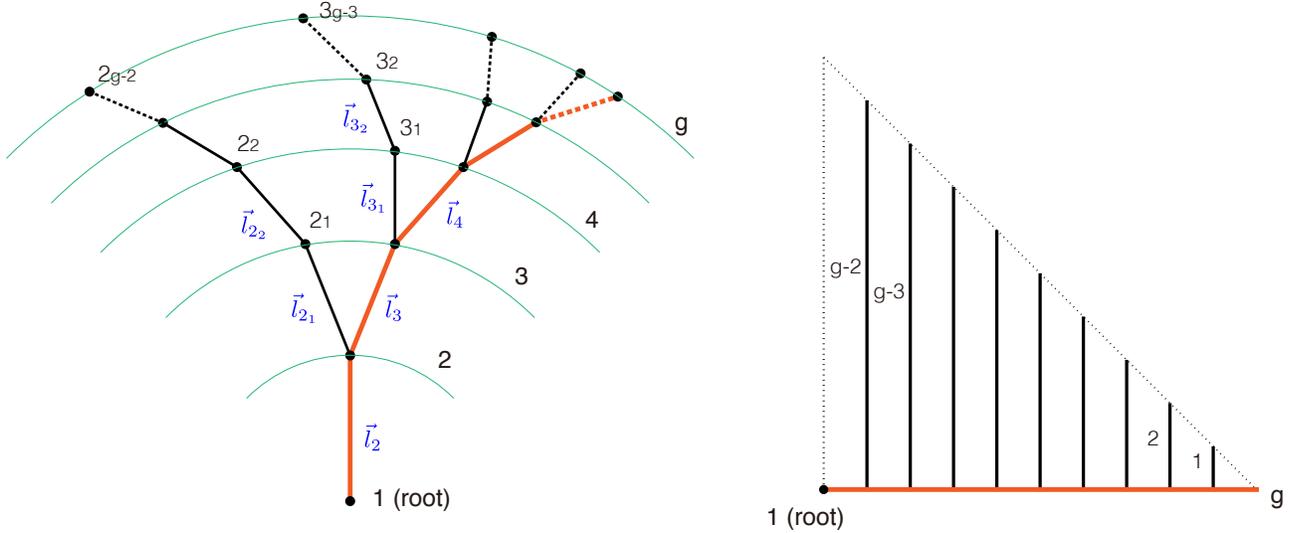}
\caption{A triangular polymer. The main backbone is indicated by the red bold-line (\BL{red}). The end monomer on each branch has the same generation number, $g$, as the main backbone.}\label{TriangularIllust}
\end{center}
\end{figure}

It has been shown experimentally\cite{Terao} that comb polymers with various lengths of side chains, equally, behave as a linear polymer. It seems obvious that, given $N\rightarrow\infty$, those findings can be generalized to all comb polymers that have finite lengths of side chains. On the basis of those findings, we examine, in this paper, a special comb polymer having long side chains that grow indefinitely in parallel with the growing backbone. One of such polymers is illustrated in Fig. \ref{TriangularIllust}; this polymer has a geometry that can be deformed to the isosceles right triangle, so that all the end monomers on the side chains have the same generation number, $g$, as the main backbone (red bold-line). So, irrespective of the size of $g$, this comb polymer preserves the same geometric structure. For this reason, we name this special polymer \textit{triangular polymer}. At first sight, such a polymer might appear to be an artificial construct. On the contrary, this is a real object that must be necessarily formed, in a certain probability, in the process of the random branching reactions under the principle of the equal reactivity of functional units (ERF)\cite{Flory}.
%%%%%%%%%%%%%%%%%% Fig. 1

\section{Theoretical}
Let the triangular polymer be constructed from $g$ generations. Then, index the branching units on the backbone from 1 (root) to $g$, while for the units on the side chains from $k_{1}$ to $k_{g-k}$, where $k_{i}$ signifies the $i$th monomer on the side chain emanating from the $k$th generation ($i=1, \cdots, g-k$); for instance, $3_{2}$ denotes the second monomers on the side chain emanating from the branching unit on the third generation (Fig. \ref{TriangularIllust}). So we consider each side chain to be a part of the corresponding monomer on the main backbone. Let $u_{g_{i}}$ represent the total number of monomers belonging to the branching unit on the $i$th generation: for instance, $u_{g_{1}}=1$, $u_{g_{2}}=1+(g-2)$, $u_{g_{3}}=1+(g-3)$, and so forth. The sum of the number of monomer units from $i=1$ to $k$ is
%%%%%%%%%%%%%%%%%% Eq. 1
\begin{equation}
u_{k}=\sum_{i=1}^{k}u_{g_{i}}=k+\frac{1}{2}(k-1)(2g-k-2)\label{COMD-1}
\end{equation}
The total number of branching units of this polymer is, therefore,
%%%%%%%%%%%%%%%%%% Eq. 2
\begin{equation}
N=u_{g}=g+\frac{1}{2}(g-1)(g-2)\label{COMD-2}
\end{equation}
To calculate the mass distribution function, we make use of the Isihara Formula\cite{Isihara, Debye}:
%%%%%%%%%%%%%%%%%% Eq. 3
\begin{equation}
\vec{r}_{Gp}=\vec{r}_{1p}-\frac{1}{N}\sum_{p=1}^{N}\vec{r}_{1p}\label{COMD-3}
\end{equation}
As we have discussed in the preceding paper\cite{Kazumi}, the end-to-end distance, $\vec{r}_{Gp}$, from the center of gravity to the $p$th monomer can be expressed by the sum of bond vectors having unequal step lengths: $\vec{r}_{Gp}=\sum_{j}c_{j}\,\vec{l}_{j}$\cite{Weiss, Redner}. So, our task is only to decompose the end-to-end vector, with the help of the Isihara formula\cite{Isihara, Debye}, into the sum of every bond vector that constitutes the polymer. Let the monomer 1 be located on the root. It is convenient to redefine $c_{k}$ as the coefficient for $\vec{l}_{k+1}$, where $\vec{l}_{k+1}=\vec{r}_{G(k+1)}-\vec{r}_{Gk}$. For the vectors on the side chains, we add the small subscript, $i$, such that $(k+1)_{i}$. Clearly, the coefficient, $c_{k}$, represents the number of trails that pass through the bond in question. According to Fig. \ref{TriangularIllust}, it follows that
%%%%%%%%%%%%%%%%%% Eq. 4
\begin{equation}
c_{k}=
\begin{cases}
N-u_{k} & (\text{for main backbone})\\[3mm]
\displaystyle\sum_{i=1}^{g-k-1}(g-k-i) & (\text{for side chains})
\end{cases}\label{COMD-4}
\end{equation}
Then the result is
%%%%%%%%%%%%%%%%%% Eq. 5
\begin{description}
\item[] for $p=1$
\begin{equation}
\vec{r}_{G1}=-\frac{1}{N}\sum_{k=1}^{g-1}\Bigg[(N-u_{k})\,\vec{l}_{k+1}+\sum_{i=1}^{g-k-1}(g-k-i)\,\vec{l}_{(k+1)_{i}}\Bigg]\label{COMD-5}
\end{equation}
 
\item[]  for $p=2$
%%%%%%%%%%%%%%%%%% Eq. 6
\begin{equation}
\vec{r}_{G2}=\frac{1}{N}\left\{\left[N-(N-u_{1})\right]\,\vec{l}_{2}-\sum_{k=1}^{g-1}\Bigg[(N-u_{k})\,\vec{l}_{k+1}+\sum_{i=1}^{g-k-1}(g-k-i)\,\vec{l}_{(k+1)_{i}}\Bigg]+(N-u_{1})\,\vec{l}_{2}\right\}\label{COMD-6}
\end{equation}
%%%%%%%%%%%%%%%%%% Eq. 7
\begin{multline}
\vec{r}_{G2_{j}}=\frac{1}{N}\Bigg\{\left[N-(N-u_{1})\right]\,\vec{l}_{2}+\sum_{i=1}^{j}[N-(g-1-i)]\,\vec{l}_{2_{i}}-\sum_{k=1}^{g-1}\Bigg[(N-u_{k})\,\vec{l}_{k+1}+\sum_{i=1}^{g-k-1}(g-k-i)\,\vec{l}_{(k+1)_{i}}\Bigg]\\
+[N-u_{1}]\,\vec{l}_{2}+\sum_{i=1}^{j}(g-1-i)\,\vec{l}_{2_{i}}\Bigg\}
\label{COMD-7}
\end{multline}
where $1\le j\le g-2$.
\item[]  for $p=3$
%%%%%%%%%%%%%%%%%% Eq. 8, 9
\begin{multline}
\vec{r}_{G3}=\frac{1}{N}\Bigg\{\sum_{k=1}^{2}\left[N-(N-u_{k})\right]\,\vec{l}_{k+1}-\sum_{k=1}^{g-1}\Bigg[(N-u_{k})\,\vec{l}_{k+1}+\sum_{i=1}^{g-k-1}(g-k-i)\,\vec{l}_{(k+1)_{i}}\Bigg]\\
+\sum_{k=1}^{2}[N-u_{k}]\,\vec{l}_{k+1}\Bigg\}\label{COMD-8}
\end{multline}
\begin{multline}
\vec{r}_{G3_{j}}=\frac{1}{N}\Bigg\{\sum_{k=1}^{2}\left[N-(N-u_{k})\right]\,\vec{l}_{k+1}+\sum_{i=1}^{j}[N-(g-2-i)]\,\vec{l}_{3_{i}}-\sum_{k=1}^{g-1}\Bigg[(N-u_{k})\,\vec{l}_{k+1}+\sum_{i=1}^{g-k-1}(g-k-i)\,\vec{l}_{(k+1)_{i}}\Bigg]\\
+\sum_{k=1}^{2}[N-u_{k}]\,\vec{l}_{k+1}+\sum_{i=1}^{j}(g-2-i)\,\vec{l}_{3_{i}}\Bigg\}
\label{COMD-9}
\end{multline}
where $1\le j\le g-3$.
\end{description}

\noindent It is obvious that we can write quite generally the end-to-end vectors from the center of gravity to the monomers on the $h$th generation in the form:
%%%%%%%%%%%%%%%%%% Eq. 10
\begin{multline}
\vec{r}_{Gh}=\frac{1}{N}\Bigg\{\sum_{k=1}^{h-1}\left[N-(N-u_{k})\right]\,\vec{l}_{k+1}-\sum_{k=1}^{g-1}\Bigg[(N-u_{k})\,\vec{l}_{k+1}+\sum_{i=1}^{g-k-1}(g-k-i)\,\vec{l}_{(k+1)_{i}}\Bigg]\\
+\sum_{k=1}^{h-1}[N-u_{k}]\,\vec{l}_{k+1}\Bigg\}\label{COMD-10}
\end{multline}
%%%%%%%%%%%%%%%%%% Eq. 11
\begin{multline}
\vec{r}_{Gh_{j}}=\frac{1}{N}\Bigg\{\sum_{k=1}^{h-1}\left[N-(N-u_{k})\right]\,\vec{l}_{k+1}+\sum_{i=1}^{j}[N-(g-h+1-i)]\,\vec{l}_{h_{i}}-\sum_{k=1}^{g-1}\Bigg[(N-u_{k})\,\vec{l}_{k+1}+\sum_{i=1}^{g-k-1}(g-k-i)\,\vec{l}_{(k+1)_{i}}\Bigg]\\
+\sum_{k=1}^{h-1}[N-u_{k}]\,\vec{l}_{k+1}+\sum_{i=1}^{j}(g-h+1-i)\,\vec{l}_{h_{i}}\Bigg\}\label{COMD-11}
\end{multline}
where $1\le h\le g$ for $\vec{r}_{Gh}$, and $2\le h\le g$ and $1\le j\le g-h$ for $\vec{r}_{Gh_{j}}$.

For the freely jointed chain, $\left\langle\vec{l}_{i}\cdot\vec{l}_{j}\right\rangle=0$ if $i\neq j$. The mean squares of the radius of gyration for respective vectors are therefore written in the form:
%%%%%%%%%%%%%%%%%% Eq. 12
\begin{equation}
\hspace{-4.0cm}\left\langle r_{Gh}^{2}\right\rangle=\frac{l^{2}}{N^{2}}\Bigg\{\sum_{k=1}^{h-1}u_{k}^{2}+\sum_{k=1}^{g-1}(N-u_{k})^{2}+\sum_{k=1}^{g-1}\sum_{i=1}^{g-k-1}(g-k-i)^{2}-\sum_{k=1}^{h-1}(N-u_{k})^{2}\Bigg\}\label{COMD-12}
\end{equation}
%%%%%%%%%%%%%%%%%% Eq. 13
\begin{multline}
\left\langle r_{Gh_{j}}^{2}\right\rangle=\frac{l^{2}}{N^{2}}\Bigg\{\sum_{k=1}^{h-1}u_{k}^{2}+\sum_{i=1}^{j}[N-(g-h+1-i)]^{2}+\sum_{k=1}^{g-1}(N-u_{k})^{2}+\sum_{k=1}^{g-1}\sum_{i=1}^{g-k-1}(g-k-i)^{2}\\
-\sum_{k=1}^{h-1}(N-u_{k})^{2}-\sum_{i=1}^{j}(g-h+1-i)^{2}\Bigg\}\label{COMD-13}
\end{multline}
where $1\le h\le g$ for $\left\langle r_{Gh}^{2}\right\rangle$, and $2\le h\le g$ and $1\le j\le g-h$ for $\left\langle r_{Gh_{j}}^{2}\right\rangle$.\\

\noindent The minus signs of the last terms in Eqs. (\ref{COMD-12}) and (\ref{COMD-13}) are necessary, because these terms are simply remainders in the arithmetic operation. As discussed in the preceding works\cite{Weiss, Redner, Kazumi}, the end-to-end vectors, $\{\vec{r}_{Gh}, \vec{r}_{Gh_{i}}\}$, with unequal step lengths should become Gaussian, if $g$ is sufficiently large. Hence, the radial mass distribution around the center of gravity can be expressed in the form:
%%%%%%%%%%%%%%%%%% Eq. 14
\begin{multline}
\varphi_{\text{triang}}(s)=\frac{1}{N}\left\{\sum_{h=1}^{g}\left(\frac{d}{2\pi\left\langle r_{Gh}^{2}\right\rangle}\right)^{\frac{d}{2}}\text{Exp}\left(-\frac{d}{2\left\langle r_{Gh}^{2}\right\rangle}s^2\right)\right.\\
\left.+\sum_{h=2}^{g}\sum_{j=1}^{g-h}\left(\frac{d}{2\pi\left\langle r_{Gh_{j}}^{2}\right\rangle}\right)^{\frac{d}{2}}\text{Exp}\left(-\frac{d}{2\left\langle r_{Gh_{j}}^{2}\right\rangle}s^2\right)\right\}\label{COMD-14}
\end{multline}
Let $S_{d}(s)$ be the surface area of the $d$-dimensional sphere having the radius, $s$. The mean square of the radius of gyration for the triangular polymer can be calculated by the equation:
%%%%%%%%%%%%%%%%%% Eq. 15
\begin{equation}
\left\langle s_{N}^{2}\right\rangle_{0}=\int_{0}^{\infty}s^{2}\varphi_{\text{triang}}(s)S_{d}(s)ds\label{COMD-15}
\end{equation}
to yield
%%%%%%%%%%%%%%%%%% Eq. 16
\begin{equation}
\left\langle s_{N}^{2}\right\rangle_{0}=\frac{g(g-1)(7g^{3}-23g^{2}+42g-18)}{15(g^{2}-g+2)^{2}}\,l^{2}\label{COMD-16}
\end{equation}
For $g\rightarrow\infty$, Eq. (\ref{COMD-16}) converges to
%%%%%%%%%%%%%%%%%% Eq. 17
\begin{equation}
\left\langle s_{N}^{2}\right\rangle_{0}\doteq\frac{7}{15}\,g\,l^{2}\label{COMD-17}
\end{equation}
The mean square of the radius of gyration is a linear function of $g$, identically to the case of dendrimers\cite{Kazumi, Yang}. By Eq. (\ref{COMD-2}), $N$ increases with $g$ as $N\propto g^{2}$, so that Eq. (\ref{COMD-17}) yields
%%%%%%%%%%%%%%%%%% Eq. 18
\begin{equation}
\left\langle s_{N}^{2}\right\rangle_{0}\propto N^{\frac{1}{2}}\,l^{2}\hspace{0.7cm}(N\rightarrow\infty)\label{COMD-18}
\end{equation}
Hence $\nu_{0}=1/4$, the same exponent as observed for the randomly branched polymer. Within our knowledge, this is probably the first example that a single pure polymer obeys the 1/4 power law.

\begin{shaded}
\vspace{-3mm}
\subsection*{Mathematical Check}
We want to check whether Eqs. (\ref{COMD-12})-(\ref{COMD-16}) are correct mathematical descriptions. Applying $g=2$ to Eq. (\ref{COMD-16}), we have the obvious result: $\left\langle s_{2}^{2}\right\rangle_{0}=\frac{1}{4}l^{2}$, while applying $g=3$, we have $\left\langle s_{N}^{2}\right\rangle_{0}=\frac{9}{16}\,l^{2}$ in agreement with the result obtained by the application of the dendrimer equations to the branched tetramer (\raisebox{-0.5mm}{\includegraphics[width=3.5mm]{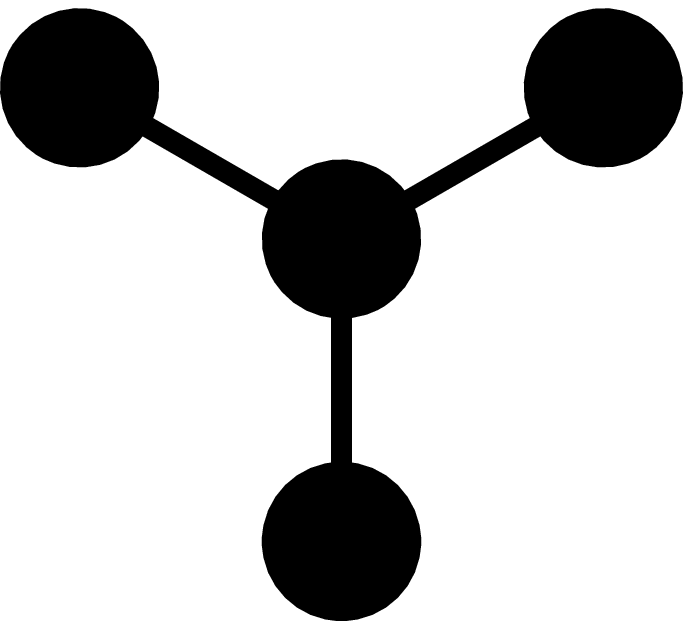}})\cite{Kazumi}. For $g=4$, Eq. (\ref{COMD-16}) gives $\left\langle s_{N}^{2}\right\rangle_{0}=\frac{46}{49}l^{2}$ in exact agreement with the calculation using the Kramers theorem\cite{Kramers}: $\left\langle s_{N}^{2}\right\rangle_{0}=\frac{l^{2}}{N^{2}}(N-1)\sum_{k=1}^{N-1}\omega_{k}\,k(N-k)$, where $\omega_{1}=\frac{4}{6}$, $\omega_{2}=\omega_{3}=\frac{1}{6}$, and $N=7$.
\end{shaded}

\section{Simulation}
\begin{figure}[H]
\begin{center}
\includegraphics[width=9.5cm]{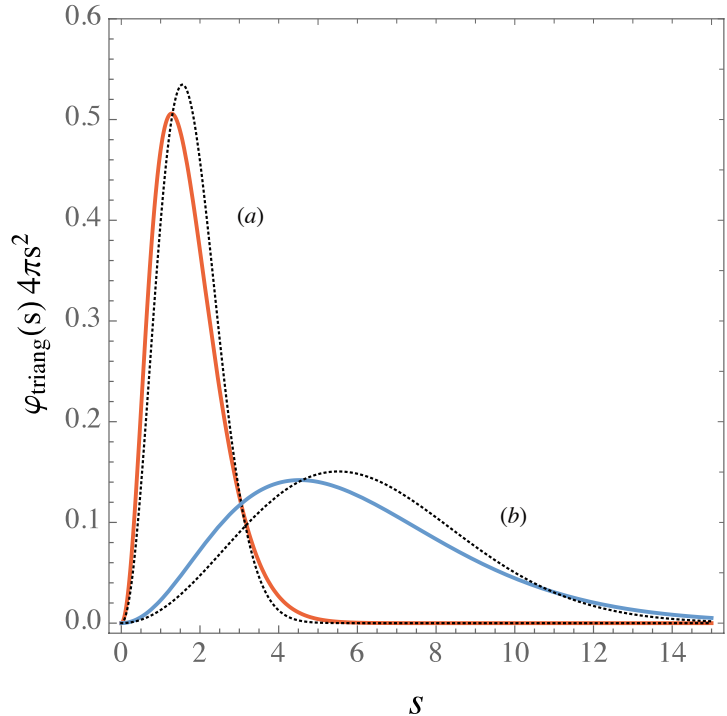}
\caption{The probability distribution of segments around the center of gravity for the triangular polymer illustrated in Fig. \ref{TriangularIllust}. The solid lines are drawn according to Eq. (\ref{COMD-14}) for (a)  $g=10$ ($N=46$) and (b) for $g=100$ ($N=4951$). The dotted lines represent the Gaussian functions having the same mean radii of gyration, $\left\langle s_{N}^{2}\right\rangle_{0}$.}\label{TriangularPDF}
\end{center}
\end{figure}

In Fig. \ref{TriangularPDF}, we plot the numerical solution of Eq. (\ref{COMD-14}), with the help of Eqs. (\ref{COMD-12})-(\ref{COMD-13}), for (a) $g=10$ ($N=46$) and (b) $g=100$ ($N=4951$). The dotted lines represent the Gaussian distribution functions having the same mean radii of gyration. There is a real difference between the exact distributions (solid lines: \BL{red} \& \BL{blue}) and the Gaussian functions ($\cdots$).

%%%%%%%%%%%%%%%%%% Fig. 2

\section{Extension and Discussion}\label{discussion}
In Table \ref{Exp-Table1}, we summarize the exponents for the ideal conformations of various polymers. To date, only three values of $\nu_{0}$ have been discovered, 1/2, 1/4, and 0, for polymeric compounds. No other exponent has been reported to date. The most prominent feature on the configurational statistics of polymers might be that the mean square of the radius of gyration can be written empirically in the form: $\left\langle s_{N}^{2}\right\rangle_{0}=A\, g\,l^{2}$ ($g\rightarrow\infty$), where $A$ is a polymer-species-dependent coefficient and also depends on the choice of the root monomer. To confirm this relationship, we have recast, in Table \ref{MeanRG}, the radii of gyration for various polymers in terms of $g$. It is seen that for all the polymers cited here, the mean squares of the radii of gyration take the form: $\left\langle s_{N}^{2}\right\rangle_{0}\propto g\,l^{2}$. 

If we accept the empirical formula, $\left\langle s_{N}^{2}\right\rangle_{0}=A\, g\,l^{2}$, as a general rule, the problem of seeking the exponent, $\nu_{0}$, reduces to the problem of seeking the relationship between $g$ and $N$.

%%%%%%%%%%%%%%%%%% Table 1
\vspace*{4mm}
 \begin{table}[h]
 \centering
  \begin{threeparttable}
    \caption{The exponents, $\nu_{0}$, for the ideal conformation, $\langle s_{N}^{2}\rangle_{0}\propto N^{2\nu_{0}}$.}\label{Exp-Table1}
\vspace*{-2mm}
  \begin{tabular}{l c}
\hline\\[-2mm]
\hspace{10mm} polymer  \hspace{30mm} &exponent $\nu_{0}$ \,\,\,\,\\[2mm]
\hline\\[-1.5mm]
\hspace{3mm} linear, star, regular comb polymers\tnote{\,a}   \hspace{10 mm} &$\frac{1}{2}$\\[1.5mm]
\hspace{3mm} isosceles triangular polymer (this work)   \hspace{10 mm} &$\frac{1}{4}$\\[1.5mm]
\hspace{3mm} randomly branched polymer\tnote{\,b}  \hspace{2 mm}(as a mixture of isomers) \hspace{10 mm} &$\frac{1}{4}$\\[1.5mm]
\hspace{3mm} dendrimers\tnote{\,c}   \hspace{10 mm} &$0$\\[1.5mm]
\hline\\[-6mm]
   \end{tabular}
   \vspace*{1mm}
   \begin{tablenotes}
     \item a. Appendix \ref{Appendix B} of this paper and the reference \cite{Terao};  b. \cite{Zimm, Dobson}; c. \cite{Kazumi, Polinska, Yang}
   \end{tablenotes}
  \end{threeparttable}
  \vspace*{-0mm}
\end{table}
%%%%%%%%%%%%%%%%%% Table 2
 \begin{table}[h]
 \centering
  \begin{threeparttable}
    \caption{The asymptotic formulae for the mean square of the radius of gyration.}\label{MeanRG}
\vspace*{-2mm}
  \begin{tabular}{l r l}
\hline\\[-2mm]
\hspace{5mm} polymer  \hspace{10mm} &formulae for $\left\langle s_{N}^{2}\right\rangle_{0}$ &\hspace{4mm} relationship between $N$ and $g$\,\,\,\,\\[2mm]
\hline\\[-1.5mm]
\hspace{5mm} linear  & $\big(\frac{1}{6}\big)\,g\, l^{2}$\hspace{7 mm} &\hspace{4mm}  $N=g$\\[1.5mm]
\hspace{5mm} star\tnote{\,a}  &$\big(\frac{3f-2}{6f}\big)\,g\, l^{2}$\hspace{7 mm} &\hspace{4mm}  $N=1+f(g-1)$\\[1.5mm]
\hspace{5mm} comb\tnote{\,b}  &$\big(\frac{1}{6}\big)\,g\, l^{2}$\hspace{7 mm} &\hspace{4mm}  $N=g+(g-1)n$\\[1.5mm]
\hspace{5mm} triangular\tnote{\,b}  &$\big(\frac{7}{15}\big)\,g\, l^{2}$\hspace{7 mm} &\hspace{4mm}  $N=g+\frac{1}{2}(g-1)(g-2)$\\[1.5mm]
\hspace{5mm} randomly branched  &\rule[2pt]{10mm}{0.5pt}\hspace{10 mm} &\hspace{6mm}\rule[2pt]{1cm}{0.5pt}\\[1.5mm]
\hspace{5mm} dendrimers\tnote{\,c}   &$g\, l^{2}$\hspace{7 mm} &\hspace{4mm}  $N=\frac{f-3+(f-1)^{g-1}}{f-2}$\\[1.5mm]
\hline\\[-6mm]
   \end{tabular}
   \vspace*{1mm}
   \begin{tablenotes}
     \item a. derived using the dendrimer equation\cite{Kazumi}; b. this work; c. \cite{Kazumi, Polinska, Yang}
   \end{tablenotes}
  \end{threeparttable}
  \vspace*{0mm}
\end{table}

In the comb and the triangular polymers, we have, by definition, one backbone. We wish to extend this architecture to a more general case and apply the empirical rule, $\left\langle s_{N}^{2}\right\rangle_{0}=A\, g\,l^{2}$, to deduce the exponent without entering the intricate vectorial calculation. We have shown, in Appendix \ref{Appendix A}, such an architecture, the extended triangular model, in which $m$ functional units (FU's) branch off, whereas $f-1-m$ FU's extend linearly. The resultant equation reads:
%%%%%%%%%%%%%%%%%% Eq. A3
\begin{equation}
N=
\begin{cases}
g+\frac{1}{2}(f-2)(g-1)(g-2) & (m=1)\\[3mm]
\displaystyle\frac{(f-2)m^{g-1}-(m-1)(f-m-1)g+(f-2)(m-2)}{(m-1)^{2}} & (m\ge 2)
\end{cases} \tag{\ref{A3}}
\end{equation}
In this extended architecture, the triangular polymer in the text corresponds to the $m=1$ case with $f=3$ to yield $g\doteq \sqrt{N}$ and $\nu_{0}=1/4$, and the dendrimer the $m=f-1$ case to yield $g\doteq constant\cdot\log N$ and $\nu_{0}=0$. For $2\le m\le f-1$, the total mass has the form: $g\propto \log N$, which necessarily gives $\nu_{0}=0$. According to the above-mentioned empirical rule, there is no other exponent than 0 and 1/4 within this extended model.

Another possible architecture may be devised. An intriguing case is in the regular comb polymer theorized in Appendix \ref{Appendix B}: This polymer has the total mass, $N=g+(g-1)n$. In the derivation in Appendix \ref{Appendix B}, we have introduced $g$ and $n$ as constants and not variables. Let us alter the length of $n$, with $g$ being fixed. Then move $g$ to infinity. Given the finite length, $n$, of the side chains, we have $N\propto g$, and the exponent should take the value, $\nu_{0}=1/2$, whereas for a large $n$ comparable to $g$, we have $N\approx g^{2}$, and the exponent takes the different value, $\nu_{0}=1/4$, the value observed for the triangular polymer (Table \ref{MeanRG}). Along this line, it might be useful to introduce the general discrete function, $n=g^{t}$ ($t\ge 0$), with $g^{t}$ representing a set of integers. What we must be careful in introducing such a general function is that it might change the mathematical form of the empirical equation. To check this possibility, let us return to the original equation of the mean square of the radius of gyration for the comb polymer (Appendix \ref{Appendix B}): 
%%%%%%%%%%%%%%%%%% B-13
\begin{equation}
\left\langle s_{N}^{2}\right\rangle_{0}=\frac{1}{6}\cdot\frac{(n+1)^{2}g^{3}+3n^{2}(n+1)g^{2}-(n+1)(8n^{2}+2n+1)g+n(n+1)(5n+1)}{(n+1)^{2}g^{2}-2n(n+1)g+n^{2}}\,l^{2}\tag{\ref{B-13}}
\end{equation}
along with the relation: $N=g+(g-1)n$. Substituting $n=g^{t}$ into Eq. (\ref{B-13}), we have, for $g\rightarrow\infty$,
%%%%%%%%%%%%%%%%%% Eq. 19
\begin{equation}
\left\langle s_{N}^{2}\right\rangle_{0}\propto
\begin{cases}
g\,l^{2} & (0\le t\le 1)\\[3mm]
g^{t}\,l^{2} & (t> 1)
\end{cases} \label{COMD-19}
\end{equation}
together with $N\doteq g^{t+1}$. For $t>1$, we have $n>g$, and therefore it will be more proper to regard $n$ rather than $g$ as the main backbone. Then, making a variable transformation, $g^{t}\Rightarrow g'$, we have $\left\langle s_{N}^{2}\right\rangle_{0}\propto g'\,l^{2}$ $ (t> 1)$, showing that the above-mentioned empirical rule is still valid. On this basis, it will be useful to redefine the empirical equation in the form:
%%%%%%%%%%%%%%%%%% Eq. 20
\begin{equation}
\left\langle s_{N}^{2}\right\rangle_{0}\propto [\text{the contour length of the main backbone}]\times l \label{COMD-20}
\end{equation}
Substituting $g=N^{\frac{1}{t+1}}$ into Eq. (\ref{COMD-19}), we have
%%%%%%%%%%%%%%%%%% Eq. 21
\begin{equation}
\nu_{0}=
\begin{cases}
\frac{1}{2(t+1)} & (0\le t\le 1)\\[3mm]
\frac{t}{2(t+1)} & (t > 1)
\end{cases} \label{COMD-21}
\end{equation}

\begin{figure}[h]
\begin{center}
\includegraphics[width=9.5cm]{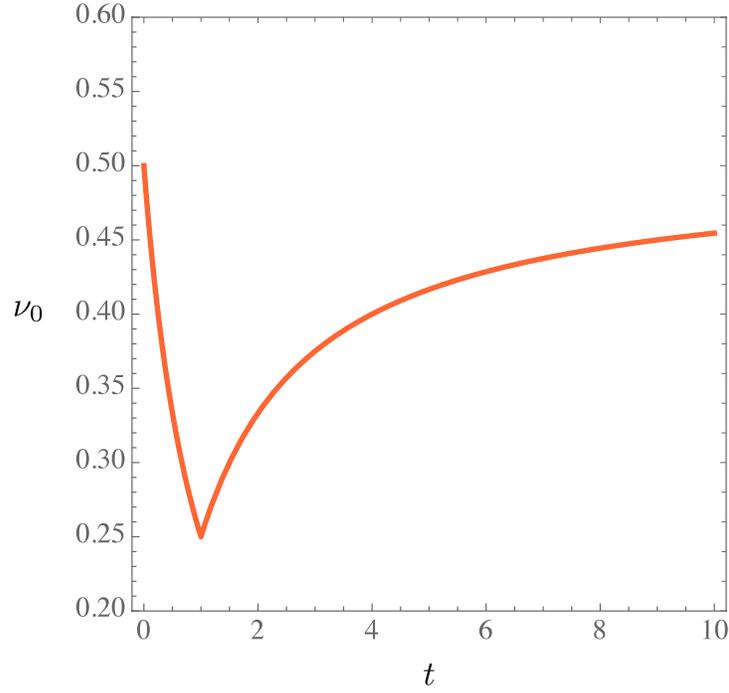}
\caption{The exponent, $\nu_{0}$, for the unperturbed conformation as a function of $t$ for the comb polymer illustrated in Fig. \ref{regularcomb}.}\label{CombExponent}
\end{center}
\end{figure}

Eq. (\ref{COMD-21}) has been plotted in Fig. \ref{CombExponent} as a function of $t$. $t=0$ corresponds to the comb polymers with the finite length of $n$ and yields $\nu_{0}=1/2$, while $t=\infty$ corresponds to the case of the star polymers and again yields the value, $\nu_{0}=1/2$\cite{Zimm, Terao, Kazumi}. As one can see from Fig. \ref{CombExponent}, between these two extremes of $t$, $\nu_{0}$ can take any values within the interval, $1/4\le \nu_{0}\le 1/2$, for a sufficiently large $g$.

%%%%%%%%%%%%%%%%%% Fig. 3

\begin{figure}[H]
\begin{center}
\includegraphics[width=16cm]{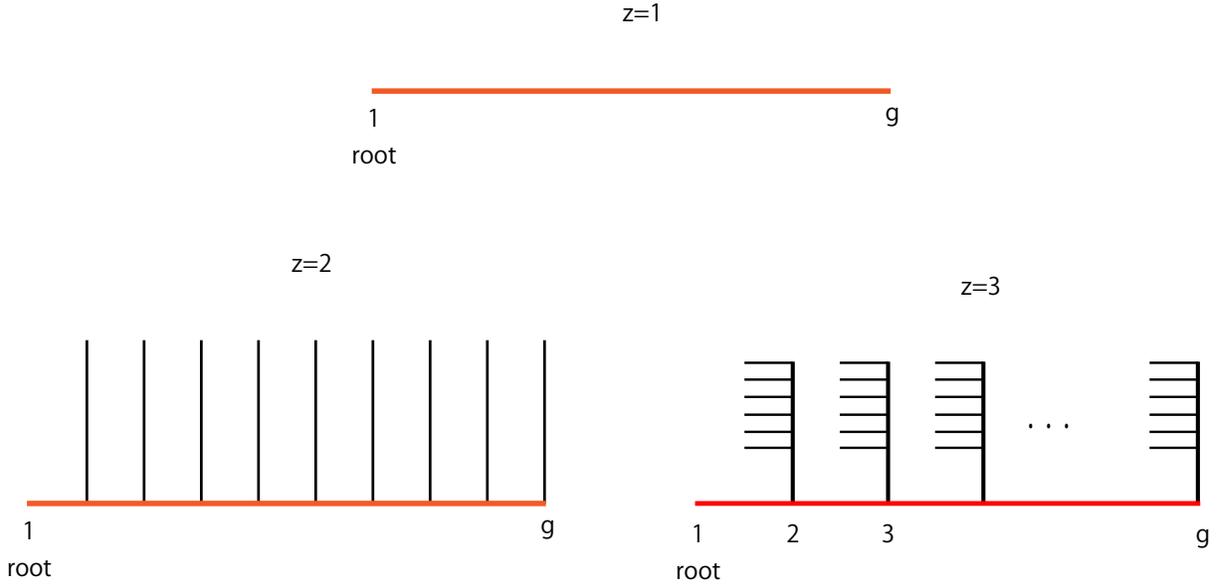}
\caption{The first ($z=1$), the second ($z=2$), and the third ($z=3$) nested structures. In $z=3$, $g-1$ comb polymers branch off from the monomers on the new backbone (red solid-line). Such nesting may be repeated successively, which results in the dendrimer structure with $f=3$.}\label{NestedStructure}
\end{center}
\end{figure}

Then let us turn our attention to the lower region, $0\le \nu_{0}\le 1/4$. Our interest is whether or not there exist architectures that allow a large number of the exponents as well in this interval. To inspect this, we introduce a nested structure. The starting polymer ($N_{1}$) may have an arbitrary architecture. In Fig. \ref{NestedStructure}, we have shown an example of $N_{1}$ being a linear polymer. The nesting rule is that on each nesting, a linear polymer with $g$ monomers is newly introduced as a backbone, on which the preceding structure is linked with each monomer on the new backbone. Such nesting is repeated successively to create deeper structures. Let $z$ represent the depth of the nest. Then, we can write generally $N_{z}=g+(g-1)N_{z-1}$. The solution to this recurrence relation is:
%%%%%%%%%%%%%%%%%% Eq. 22
\begin{equation}
N_{z}=(g-1)^{z-1}N_{1}+g\frac{(g-1)^{z-1}-1}{g-2}\hspace{5mm} (z\ge 1)\label{COMD-22}
\end{equation}
where $N_{1}$ can be any structures as mentioned above; for instance, $N_{1}=1$, $N_{1}=g$, $N_{1}(comb)=g+(g-1)n$, $N_{1}(triang)=g+\frac{1}{2}(g-1)(g-2)$, or so forth (Table \ref{MeanRG}).

\vspace{0mm}
%%%%%%%%%%%%%%%%%% Fig. 4

Here we consider the case that $N_{1}$ is a linear polymer. Substituting $N_{1}=g$ into Eq. (\ref{COMD-22}) and taking the limit, $g\rightarrow\infty$, we have $N_{z}\propto g^z$. Then applying the empirical rule (\ref{COMD-20}), we have $\left\langle s_{N}^{2}\right\rangle_{0}\propto (zgl)l$, so that for a fixed $z$, the mean square of the radius of gyration varies as $\left\langle s_{N}^{2}\right\rangle_{0}\propto zN^{\frac{1}{z}}l^{2}$, as $g\rightarrow\infty$. Hence
%%%%%%%%%%%%%%%%%% Eq. 23
\begin{equation}
\nu_{0}=\frac{1}{2\,z}\hspace{5mm} (z=1, 2, 3, \cdots)\label{COMD-23}
\end{equation}
By substituting $z=1$, we recover $\nu_{0}=1/2$, the familiar value for the linear polymer; substituting $z=2$, we recover $\nu_{0}=1/4$, the value for the special comb polymers having $n=g$; for a large $z$ limit, Eq. (\ref{COMD-23}) approaches the exponent, $\nu_{0}=0$, corresponding to the dendrimer as an ultimate structure. It is seen that, given the empirical rule (\ref{COMD-20}), we can expect an infinite number of the exponents \big($\nu_{0}=\frac{1}{2},\, \frac{1}{4},\, \frac{1}{6},\, \cdots$\big) in the interval, $0\le\nu_{0}\le 1/2$, corresponding to respective architectures from the linear polymer to the dendrimer.

\begin{shaded}
\vspace{-4mm}
\subsection*{Ultimate Structure in the Large $z$ Limit}
It seems obvious that, whatever $N_{1}$ is, the nested structure ultimately approaches the dendrimer with $f=3$ as $z\rightarrow\infty$.
\end{shaded}

The question may be raised about whether the architectures mentioned above (the comb polymers having side chains of the length, $g^{t}$, and the nesting structures) have a reality. Whether a polymer has a reality depends on the availability of the polymer through synthetic chemistry. In the sense that the above architectures can not be obtained, with a practical yield, by using the conventional random reaction, they will not be realistic. Nevertheless, not all these polymers are imaginal ones: from the thermodynamic point of view, the architectures having $\nu_{0}\ge 1/3$ can exist as it is, and some portion of the polymers having $\nu_{0}< 1/3$ can exist in the expanded state, and hence, a considerable portion of the above polymers, in principle, can be constructed by step-by-step synthesis, given sufficient time and technology. 

A noteworthy feature might be that, according to the theory of the excluded volume effects\cite{Issacson, Seitz, Kazumi, Ferber, Blavatska, Haydukivska} \big[for instance, $\nu=\frac{2(1+\nu_{0})}{d+2}$ ($d=3$)\footnote{\, We confine our discussion to $d=3$ since this formula has not been fully verified experimentally for the dimensions other than 3.}\big], the present results imply the existence of a large number of the real exponents, $\nu$, corresponding to the respective ideal exponents, $\nu_{0}$.

\section{Concluding Remarks}
The probability distribution of segments for the triangular polymer with a large $g$ can be expressed as the sum of the Gaussian functions having the mean squares of the end-to-end distances, $\left\{\left\langle r_{Gh}^{2}\right\rangle, \left\langle r_{Gh_{j}}^{2}\right\rangle\right\}$. It is found from Fig. \ref{TriangularPDF} that there is a substantial deviation of the resulting formula (\ref{COMD-14}) from the Gaussian function having the same mean square of the radius of gyration. On the other hand, in common with all the known polymers, the triangular polymer obeys the relationship:
%%%%%%%%%%%%%%%%%%
\begin{equation}
\left\langle s_{N}^{2}\right\rangle_{0}\propto \big[g\,l\big]\,l\tag{\ref{COMD-20}$'$}
\end{equation}
where $g\,l$ is the contour length of the main backbone from the root to the youngest (outermost) generation, $g$.

As a reference polymer, we have calculated the probability distribution function (PDF) of segments for the regular comb polymer (Appendix \ref{Appendix B}). As expected, the PDF of this polymer was found to be closer to the Gaussian function than that of the triangular polymer. 

By virtue of the relationship (\ref{COMD-20}$'$), we might have gained a means to deduce the exponent, $\nu_{0}$, making use of the relationship between $N$ and $g$, without entering the intricate vectorial calculations. As an example, we have put forth, in Section \ref{discussion}, an extended comb polymer that has the side chains of the length, $n=g^{t}$, along with the total mass, $N\doteq g^{t+1}$ ($g\rightarrow\infty$). This leads us to the consequence: $\nu_{0}$ can take an arbitrary value in the interval, $1/4\le\nu_{0}\le 1/2$, by changing $t$ properly if $g$ is sufficiently large. We have put forth another example of the nested structures that create a continuous change of architectures from the linear chain molecule to the dendrimer. It is shown that, according to the empirical relationship (\ref{COMD-20}$'$), those new architectures give rise to a large number of the ideal exponents, $\nu_{0}=\frac{1}{2},\, \frac{1}{4},\, \frac{1}{6},\, \cdots$, in the interval, $0\le\nu_{0}\le 1/2$.

\clearpage
%%%%%%%%%%%%%%%%%% Appendix
\renewcommand{\appendixname}{Appendix}

\appendix

\section*{\LARGE \textbf{Appendix}}
\vspace*{0.5cm}
\numberwithin{equation}{section}
\setcounter{equation}{0}
\renewcommand{\thefigure}{A\arabic{figure}}
\setcounter{figure}{0}

\section{Extended Triangular Polymers}\label{Appendix A}
%%%%%%%%%%%%%%%%%% Fig. A1
\begin{figure}[h]
\begin{center}
\includegraphics[width=8cm]{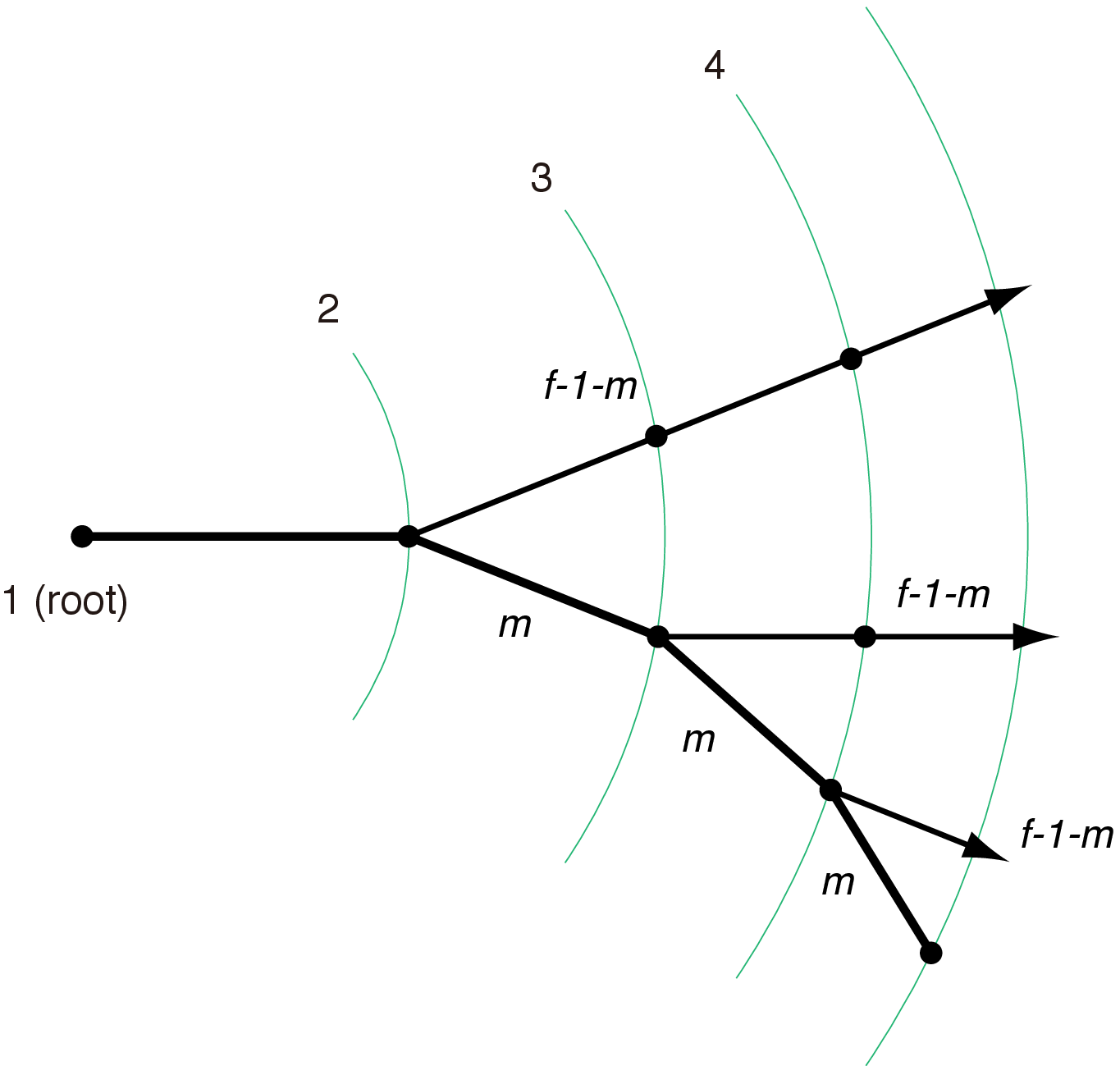}
\caption{Extended triangular model.}\label{GeneralizedTriangular-A}
\end{center}
\end{figure}
We extend the architecture of the triangular polymer to a more general case. Let us consider a polymer, in which $m$ functional units (FU's) branch off, whereas $f-1-m$ FU's extend linearly. Let every end monomer reach the $g$th generation. The triangular polymer in the text corresponds to the case of $m=1$ with $f=3$, and the dendrimer the $m=f-1$ case. The number of monomers, $u_{g_{i}}$, in the $i$th generation is
%%%%%%%%%%%%%%%%%% Eq. A1
\begin{equation}
u_{g_{1}}=1;\hspace{2mm} u_{g_{i}}=m^{i-2}[1+(f-1-m)(g-i)] \hspace{0.5cm}(i\ge 2) \label{A1}
\end{equation}
Hence the total mass of this polymer is
%%%%%%%%%%%%%%%%%% Eq. A2
\begin{equation}
N=1+\sum_{i=2}^{g}m^{i-2}[1+(f-1-m)(g-i)] \hspace{0.5cm} (m\ge 2) \label{A2}
\end{equation}
which gives
%%%%%%%%%%%%%%%%%% Eq. A3
\begin{equation}
N=
\begin{cases}
g+\frac{1}{2}(f-2)(g-1)(g-2) & (m=1)\\[3mm]
\displaystyle\frac{(f-2)m^{g-1}-(m-1)(f-m-1)g+(f-2)(m-2)}{(m-1)^{2}} & (m\ge 2)
\end{cases} \label{A3}
\end{equation}
For $m=1$ and $f=3$, this is equivalent to Eq. (\ref{COMD-2}) in the text, so that we have $g\propto \sqrt{N}$ giving $\nu_{0}=1/4$, whereas, for $m\ge 2$, we have $g\propto \log N$ giving $\nu_{0}=0$, the value identical to the dendrimer case\cite{Kazumi}.

%\clearpage
%\vspace*{-0.5cm}
%%%%%%%%%%%%%%%%%%
\numberwithin{equation}{section}
\setcounter{equation}{0}
\renewcommand{\thefigure}{B\arabic{figure}}
\setcounter{figure}{0}

\vspace*{1cm}
\section{Regular Comb Polymers}\label{Appendix B}
\begin{figure}[h]
\begin{center}
\includegraphics[width=18cm]{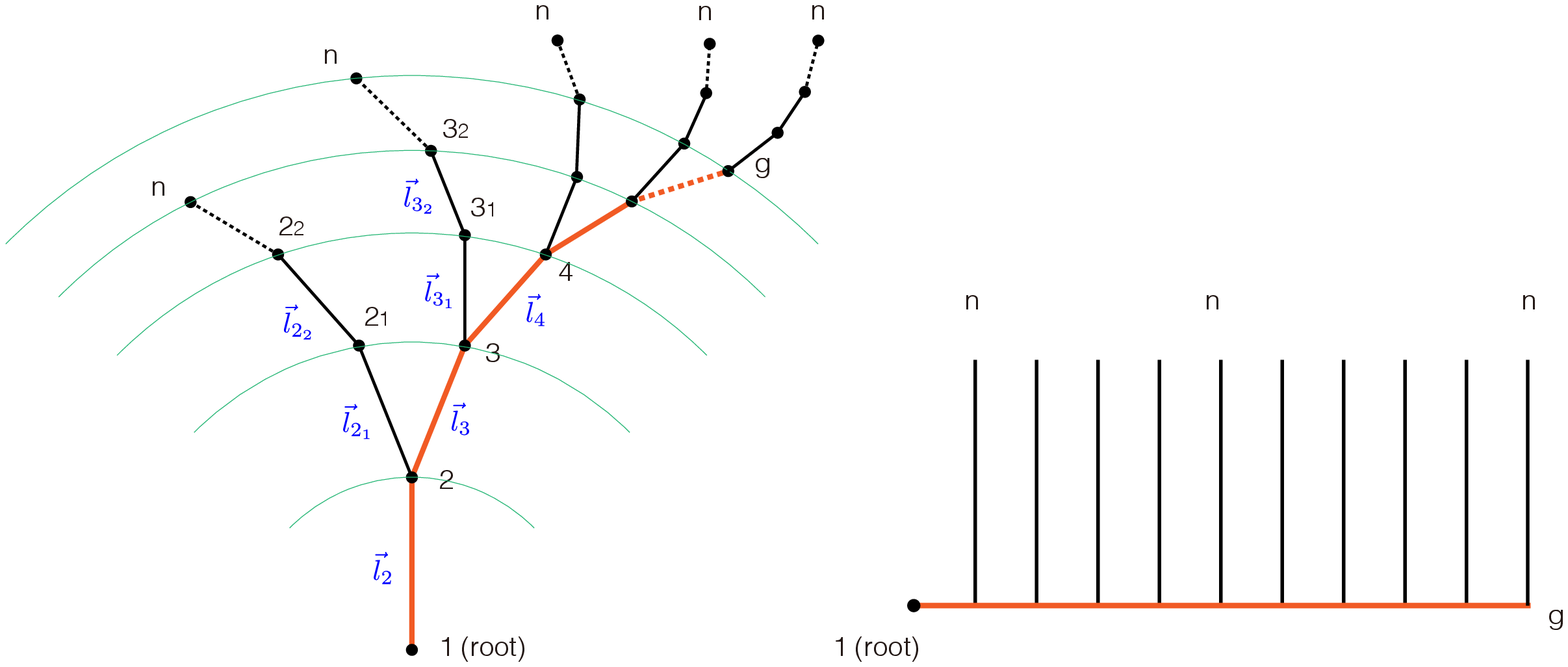}
\caption{The regular comb polymer having the same length, $n$, of side chains. The red bold-line is the main backbone, and the black lines are side chains.}\label{regularcomb}
\end{center}
\end{figure}

In this section, we calculate the probability distribution function (PDF) for the regular comb polymers (Fig. \ref{regularcomb}). Let the main backbone of a comb polymer be made up $g$ generations with $g-1$ side chains of the length, $n$. So, this polymer has the molecular mass: $N=g+(g-1)n$. Let us index the branching units on the backbone from 1 to $g$, and the units on the side chains from 1 to $n$; for instance, $3_{2}$ denotes the second monomers on the side chain emanating from the branching unit of the third generation (Fig. \ref{regularcomb}).

%%%%%%%%%%%%%%%%%% Fig. B1

Following the definition in the text, let $u_{k}$ be the sum of the number of monomer units from the root to the $k$th generation. We have then $u_{k}=k+(k-1)n$. The end-to-end vectors, $\vec{r}_{Gp}$, are
%%%%%%%%%%%%%%%%%% B-1
\begin{description}
\item[] for $p=1$
\begin{equation}
\vec{r}_{G1}=-\frac{1}{N}\left\{\sum_{k=1}^{g-1}\left[N-k-(k-1)n\right]\,\vec{l}_{(k+1)}+\sum_{k=1}^{g-1}\sum_{i=1}^{n}(n-i+1)\,\vec{l}_{(k+1)_{i}}\right\}\label{B-1}
\end{equation}
where the small subscript $i$ in $\vec{l}_{k_{i}}$ denotes the $i$th monomer on the side chain emanating from the $k$th generation as stated above, and $\vec{l}_{(k+1)_{i}}=\vec{r}_{G(k+1)_{i+1}}-\vec{r}_{Gk_{i}}$.
\item[]  for $p=2$
%%%%%%%%%%%%%%%%%% B-2
\begin{multline}
\vec{r}_{G2}=\frac{1}{N}\left\{[N-(N-1)]\,\vec{l}_{2}-\sum_{k=1}^{g-1}\left[N-k-(k-1)n\right]\,\vec{l}_{(k+1)}\right.\\
\left.-\sum_{k=1}^{g-1}\sum_{i=1}^{n}(n-i+1)\,\vec{l}_{(k+1)_{i}}+(N-1)\,\vec{l}_{2}\right\}\label{B-2}
\end{multline}
%%%%%%%%%%%%%%%%%% B-3
\begin{multline}
\vec{r}_{G2_{j}}=\frac{1}{N}\left\{[N-(N-1)]\,\vec{l}_{2}+\sum_{i=1}^{j}[N-n+i-1]\,\vec{l}_{2_{i}}-\sum_{k=1}^{g-1}\left[N-k-(k-1)n\right]\,\vec{l}_{(k+1)}\right.\\
\left.-\sum_{k=1}^{g-1}\sum_{i=1}^{n}(n-i+1)\,\vec{l}_{(k+1)_{i}}+(N-1)\,\vec{l}_{2}+\sum_{i=1}^{j}[n-i+1]\,\vec{l}_{2_{i}}\right\}\label{B-3}
\end{multline}

\item[]  for $p=3$
%%%%%%%%%%%%%%%%%% B-4
\begin{multline}
\vec{r}_{G3}=\frac{1}{N}\left\{\sum_{k=1}^{2}[N-(N-k-(k-1)n)]\,\vec{l}_{(k+1)}-\sum_{k=1}^{g-1}\left[N-k-(k-1)n\right]\,\vec{l}_{(k+1)}\right.\\
\left.-\sum_{k=1}^{g-1}\sum_{i=1}^{n}(n-i+1)\,\vec{l}_{(k+1)_{i}}+\sum_{k=1}^{2}[N-k-(k-1)n]\,\vec{l}_{(k+1)}\right\}\label{B-4}
\end{multline}
%%%%%%%%%%%%%%%%%% B-5
\begin{multline}
\hspace{-0.5cm}\vec{r}_{G3_{j}}=\frac{1}{N}\left\{\sum_{k=1}^{2}[N-(N-k-(k-1)n)]\,\vec{l}_{(k+1)}+\sum_{i=1}^{j}[N-n+i-1]\,\vec{l}_{3_{i}}-\sum_{k=1}^{g-1}\left[N-k-(k-1)n\right]\,\vec{l}_{(k+1)}\right.\\
\left.-\sum_{k=1}^{g-1}\sum_{i=1}^{n}(n-i+1)\,\vec{l}_{(k+1)_{i}}+\sum_{k=1}^{2}[N-k-(k-1)n]\,\vec{l}_{(k+1)}+\sum_{i=1}^{j}[n-i+1]\,\vec{l}_{3_{i}}\right\}\label{B-5}
\end{multline}

\item[]  Obviously, we can write, quite generally, as
\item[]  for $p=h$
%%%%%%%%%%%%%%%%%% B-6
\begin{multline}
\vec{r}_{Gh}=\frac{1}{N}\left\{\sum_{k=1}^{h-1}[N-(N-k-(k-1)n)]\,\vec{l}_{(k+1)}-\sum_{k=1}^{g-1}\left[N-k-(k-1)n\right]\,\vec{l}_{(k+1)}\right.\\
\left.-\sum_{k=1}^{g-1}\sum_{i=1}^{n}(n-i+1)\,\vec{l}_{(k+1)_{i}}+\sum_{k=1}^{h-1}[N-k-(k-1)n]\,\vec{l}_{(k+1)}\right\}\label{B-6}
\end{multline}
%%%%%%%%%%%%%%%%%% B-7
\begin{multline}
\hspace{-0.5cm}\vec{r}_{Gh_{j}}=\frac{1}{N}\left\{\sum_{k=1}^{h-1}[N-(N-k-(k-1)n)]\,\vec{l}_{(k+1)}+\sum_{i=1}^{j}[N-n+i-1]\,\vec{l}_{h_{i}}-\sum_{k=1}^{g-1}\left[N-k-(k-1)n\right]\,\vec{l}_{(k+1)}\right.\\
\left.-\sum_{k=1}^{g-1}\sum_{i=1}^{n}(n-i+1)\,\vec{l}_{(k+1)_{i}}+\sum_{k=1}^{h-1}[N-k-(k-1)n]\,\vec{l}_{(k+1)}+\sum_{i=1}^{j}[n-i+1]\,\vec{l}_{h_{i}}\right\}\label{B-7}
\end{multline}
where $1\le h\le g$ for $\vec{r}_{Gh}$, and  $2\le h\le g$ and $1\le j\le n$ for $\vec{r}_{Gh_{j}}$.

\end{description}

The mean squares of the end-to-end distances are
%%%%%%%%%%%%%%%%%% B-8
\begin{multline}
\left\langle r_{Gh}^{2}\right\rangle=\frac{l^{2}}{N^{2}}\left\{\sum_{k=1}^{h-1}[k+(k-1)n]^{2}+\sum_{k=1}^{g-1}\left[N-k-(k-1)n\right]^2\right.\\
\left.+(g-1)\sum_{i=1}^{n}(n-i+1)^{2}-\sum_{k=1}^{h-1}[N-k-(k-1)n]^{2}\right\}\label{B-8}
\end{multline}
%%%%%%%%%%%%%%%%%% B-9
\begin{multline}
\left\langle r_{Gh_{j}}^{2}\right\rangle=\frac{l^{2}}{N^{2}}\left\{\sum_{k=1}^{h-1}[k+(k-1)n]^{2}+\sum_{i=1}^{j}[N-n+i-1]^{2}+\sum_{k=1}^{g-1}\left[N-k-(k-1)n\right]^{2}\right.\\
\left.+(g-1)\sum_{i=1}^{n}(n-i+1)^{2}-\sum_{k=1}^{h-1}[N-k-(k-1)n]^{2}-\sum_{i=1}^{j}[n-i+1]^{2}\right\}\label{B-9}
\end{multline}
where $1\le h\le g$ for $\left\langle r_{Gh}^{2}\right\rangle$, and  $2\le h\le g$ and $1\le j\le n$ for $\left\langle r_{Gh_{j}}^{2}\right\rangle$.
We are ready to calculate the segment distribution of the comb polymers with a large $g$:
%%%%%%%%%%%%%%%%%% B-10
\begin{multline}
\varphi_{\text{comb}}(s)=\frac{1}{N}\left\{\sum_{h=1}^{g}\left(\frac{d}{2\pi\left\langle r_{Gh}^{2}\right\rangle}\right)^{\frac{d}{2}}\text{Exp}\left(-\frac{d}{2\left\langle r_{Gh}^{2}\right\rangle}s^2\right)\right.\\
\left.+\sum_{h=2}^{g}\sum_{j=1}^{n}\left(\frac{d}{2\pi\left\langle r_{Gh_{j}}^{2}\right\rangle}\right)^{\frac{d}{2}}\text{Exp}\left(-\frac{d}{2\left\langle r_{Gh_{j}}^{2}\right\rangle}s^2\right)\right\}\label{B-10}
\end{multline}
The mean square of the radius of gyration is calculated by the equation:
%%%%%%%%%%%%%%%%%% B-11
\begin{equation}
\left\langle s_{N}^{2}\right\rangle_{0}=\int_{0}^{\infty}s^{2}\varphi_{\text{comb}}(s)S_{d}(s)ds\label{B-11}
\end{equation}
to yield
%%%%%%%%%%%%%%%%%% B-12
\begin{equation}
\left\langle s_{N}^{2}\right\rangle_{0}=\frac{1}{N}\left(\sum_{h=1}^{g}\left\langle r_{Gh}^{2}\right\rangle+\sum_{h=2}^{g}\sum_{j=1}^{n}\left\langle r_{Gh_{j}}^{2}\right\rangle\right)\label{B-12}
\end{equation}
Substituting Eqs. (\ref{B-8}) and (\ref{B-9}), we have the expression of $\left\langle s_{N}^{2}\right\rangle_{0}$:
%%%%%%%%%%%%%%%%%% B-13
\begin{equation}
\left\langle s_{N}^{2}\right\rangle_{0}=\frac{1}{6}\cdot\frac{(n+1)^{2}g^{3}+\mathscr{R}_{numer}}{(n+1)^{2}g^{2}+\mathscr{R}_{denom}}\,l^{2}\label{B-13}
\end{equation}
where
%%%%%%%%%%%%%%%%%%
\begin{align}
\mathscr{R}_{numer}&=3n^{2}(n+1)g^{2}-(n+1)(8n^{2}+2n+1)g+n(n+1)(5n+1)\notag\\
\mathscr{R}_{denom}&=-2n(n+1)g+n^{2}\notag
\end{align}
For $n=0$, we recover the well-established result, $\left\langle s_{N}^{2}\right\rangle_{0}=\frac{1}{6}(N-1/N)\,l^{2}$, for linear chains. If $g\rightarrow\infty$, Eq. (\ref{B-13}) reduces to
%%%%%%%%%%%%%%%%%% B-14
\begin{equation}
\left\langle s_{N}^{2}\right\rangle_{0}\doteq\frac{1}{6}\,g\, l^{2}\label{B-14}
\end{equation}
Since $N=g+(g-1)n$, this gives $\left\langle s_{N}^{2}\right\rangle_{0}\propto N$ and $\nu_{0}=1/2$. For a finite length of $n$, a comb polymer behaves as if a linear chain, in agreement with the experimental observations\cite{Terao}. If the side chains are sufficiently large and comparable to $g$, then we have $g\propto\sqrt{N}$ and $\nu_{0}=1/4$, in accord with the case of the triangular polymer in the text.

\begin{shaded}
\vspace{-3mm}
\subsection*{Mathematical Check}
Apply $g=2$ and $N=n+2$ to Eq. (\ref{B-13}), and we recover the known result: $\left\langle s_{N}^{2}\right\rangle_{0}=\frac{1}{6}\left(N-1/N\right)$. The comb polymer with $g=3$ can be equated with the star polymer having $f=3$ and $(N_{1}, N_{2}, N_{3})=(2, n, n+1)$. Substituting $g=3$ into Eq. (\ref{B-13}), we have $\left\langle s_{N}^{2}\right\rangle_{0}=\frac{(4n^{2}+11n+12)(n+1)}{3(2n+3)^{2}}\,l^{2}$ which is exactly the same result as calculated by the star polymer equation derived in the preceding paper\cite{Kazumi}. The same result can be obtained also by the use of the Kramers theorem: $\left\langle s_{N}^{2}\right\rangle_{0}=\frac{l^{2}}{N^{2}}(N-1)\sum_{k=1}^{N-1}\omega_{k}\,k(N-k)$, where $\omega_{1}=\frac{3}{2n+2}$, $\omega_{k}=\frac{2}{2n+2}$ ($k=2, 3, \cdots, n$), $\omega_{n+1}=\frac{1}{2n+2}$, and $N=2n+3$.\textcolor{blue}{/\hspace{-0.7mm}/}
\end{shaded}

\begin{figure}[H]
\begin{center}
\includegraphics[width=9.5cm]{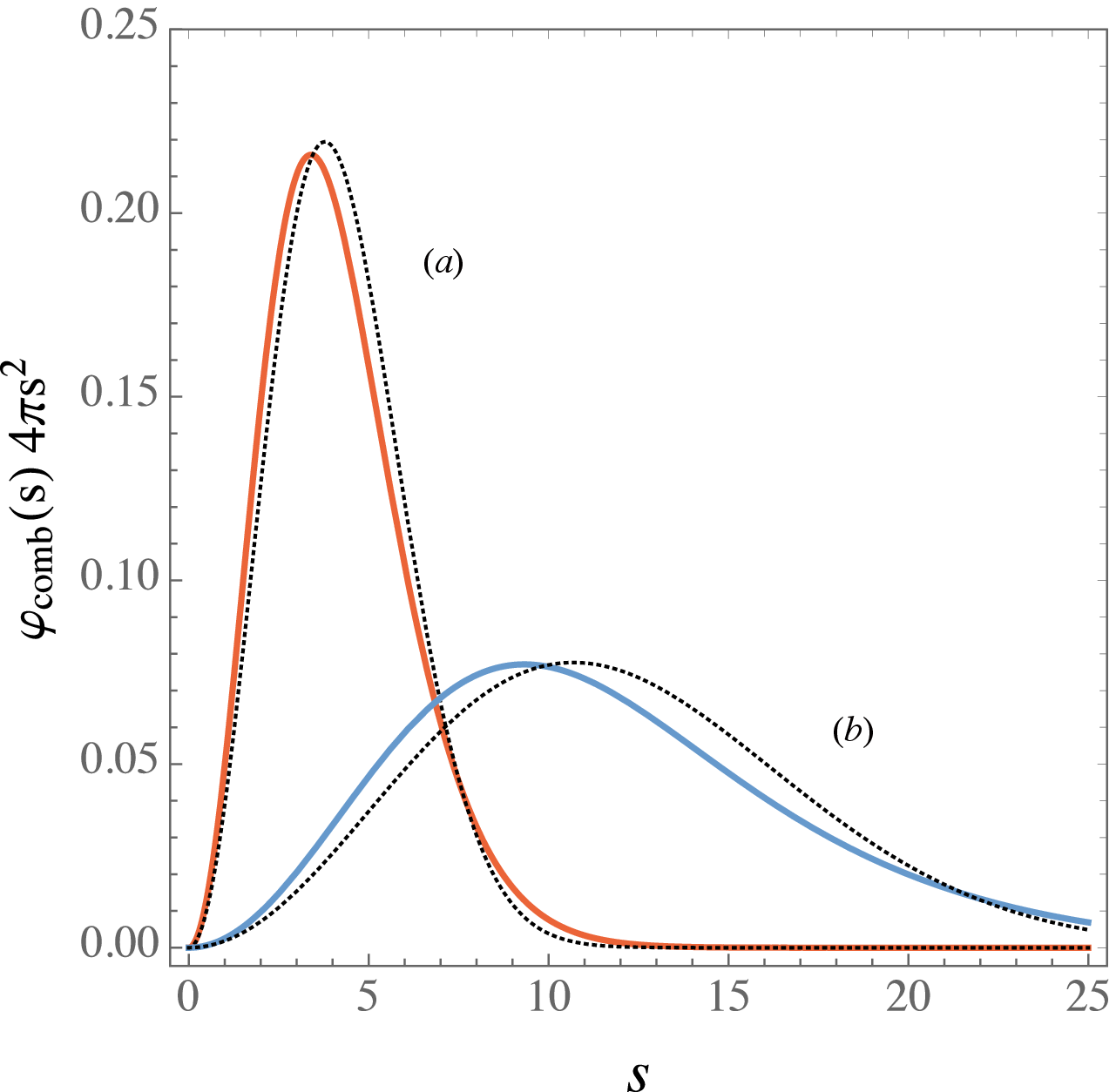}
\caption{The radial segment distributions around the center of gravity for the regular comb polymers. The red solid-line is the PDF of the comb polymer having (a) $g=100$ and $n=10$ ($N=2190$) and the blue solid-line is that for (b) $g=1000$ and $n=10$ ($N=10990$); the dotted lines are the corresponding Gaussian PDF's having the same $\left\langle s_{N}^{2}\right\rangle_{0}$'s.}\label{RegularCombSimu}
\end{center}
\end{figure}

In Fig. \ref{RegularCombSimu}, two examples of (a) $g=100$, $n=10$, and (b) $g=1000$, $n=10$ ($d=3$) are plotted according to Eq. (\ref{B-10}) with the help of Eqs. (\ref{B-8}) and (\ref{B-9}); the dotted lines show the corresponding Gaussian functions having the same $\left\langle s_{N}^{2}\right\rangle_{0}$'s. Although there is a clear difference between Eq. (\ref{B-10}) and the Gaussian PDF, the difference is smaller compared with the case of the triangular polymer (Fig. \ref{TriangularPDF}). The observed deviation is just as we had expected from Ishihara's work\cite{Isihara, Debye}: In contrast to the end-to-end distance distribution, the segment distribution around the center of gravity is, in general, not Gaussian. It is important to note that the observed feature of the deviation in Fig. \ref{RegularCombSimu} is close to the case in linear polymers\cite{Isihara, Debye}. From this example, we can infer that the segment distribution of the comb polymer having finite $n$ should rigorously approach that of the linear polymer in the limit of a large $g\gg n$.
%%%%%%%%%%%%%%%%%% Fig. B2

\clearpage
%%%%%%%%%%%%%%%%%%

\end{document}